\documentclass[journal,twoside,web]{ieeecolor}
\usepackage{tmi}
\usepackage{cite}
\usepackage{amsmath,amssymb,amsfonts}
\usepackage{caption}
\usepackage{subcaption}
\usepackage{hyperref}
\usepackage{soul}
\usepackage[ruled,vlined,linesnumbered]{algorithm2e}


\usepackage{bm}
\usepackage{graphicx}
\usepackage{textcomp}

\def\BibTeX{{\rm B\kern-.05em{\sc i\kern-.025em b}\kern-.08em
    T\kern-.1667em\lower.7ex\hbox{E}\kern-.125emX}}
\newcommand{\xrm}{\mathrm{x}}
\newcommand{\frm}{\mathrm{f}}
\newcommand{\vrm}{\mathrm{v}}
\newcommand{\xbf}{\mathbf{x}}

\newcommand{\Tbf}{\mathbf{T}}
\newcommand{\Ibf}{\mathbf{I}}

\newcommand{\Fbf}{\mathbf{F}}
\newcommand{\Mbf}{\mathbf{M}}
\newcommand{\Ubf}{\mathbf{U}}

\newcommand{\Cbf}{\mathbf{C}}

\newcommand{\RR}{\mathbb{R}}

\newcommand{\CC}{\mathbb{C}}

\DeclareMathOperator*{\argmin}{arg\,min}

\newcommand{\F}{\mathcal{F}}

\newcommand{\G}{\mathcal{G}}
\newcommand{\R}{\mathcal{R}}

\newcommand{\HH}{\mathcal{H}}
\newcommand{\D}{\mathcal{D}}

\newcommand{\Scal}{\mathcal{S}}

\begin{document}
\title{Magnetic Resonance Parameter Mapping using Self-supervised Deep Learning with Model Reinforcement}
\author{Wanyu Bian, Albert Jang, and Fang Liu
\thanks{This work was supported in part by the grants R21EB031185; R01AR081344; R01AR079442.'' }
\thanks{W. Bian, A. Jang and F. Liu are with the
Athinoula A. Martinos Center for Biomedical Imaging, Massachusetts General Hospital
and Harvard Medical School, Charlestown, MA 02129 USA. (corresponding author: F. Liu, email:fliu12@mgh.harvard.edu)}
}

\maketitle

\begin{abstract}
This paper proposes a novel self-supervised learning method, RELAX-MORE, for quantitative MRI (qMRI) reconstruction.  The proposed method uses an optimization algorithm to unroll a model-based qMRI reconstruction into a deep learning framework, enabling the generation of highly accurate and robust MR parameter maps at imaging acceleration. Unlike conventional deep learning methods requiring a large amount of training data, RELAX-MORE is a subject-specific method that can be trained on single-subject data through self-supervised learning, making it accessible and practically applicable to many qMRI studies. Using the quantitative $T_1$ mapping as an example at different brain, knee and phantom experiments, the proposed method demonstrates excellent performance in reconstructing MR parameters, correcting imaging artifacts, removing noises, and recovering image features at imperfect imaging conditions. Compared with other state-of-the-art conventional and deep learning methods, RELAX-MORE significantly improves efficiency, accuracy, robustness, and generalizability for rapid MR parameter mapping. This work demonstrates the feasibility of a new self-supervised learning method for rapid MR parameter mapping, with great potential to enhance the clinical translation of qMRI.
\end{abstract}


\section{Introduction}
\label{sec:introduction}
\IEEEPARstart{D}{eep} learning has had a profound impact on medical imaging, including MRI. The breadth of its impact includes disease diagnosis, prognosis, image analysis, and processing \cite{litjens2017survey,shen2017deep}. In particular, deep learning methods for MRI reconstruction are becoming increasingly popular due to their capability to learn important image features directly from large datasets \cite{knoll2020deep}. One popular method is estimating unacquired k-space data through training end-to-end convolutional neural networks (CNNs), using undersampled data as input and corresponding fully sampled data as a reference \cite{hammernik2018learning,schlemper2017deep,han2018deep,zhu2018image,yang2017dagan,lee2018deep,liu2019santis}. This enables the networks to learn the mapping function between the undersampled and fully sampled k-space data pairs. A well-trained network can then reconstruct undersampled k-space data for accelerated MRI. 

Despite recent advances  in deep learning MRI reconstruction \cite{hammernik2018learning,schlemper2017deep,han2018deep,zhu2018image,yang2017dagan,lee2018deep,knoll2020deep,liu2019santis,yaman2020self,blumenthal2022nlinv}, several challenges remain, including limited training data access. Acquiring fully sampled k-space data can be time-consuming and expensive, and the data collection challenge for quantitative MRI is even more pronounced. Quantitative MRI is an advanced method to quantify tissue MR parameters by modeling the acquired MR signal. For example, a popular qMRI method called variable Flip Angle (vFA) \cite{wang1987optimizing} to quantify tissue spin-lattice relaxation time ($T_1$) requires acquisition for multiple repeated scans at different flip angles. Since each acquisition can take several minutes, the repeated scan time can be non-trivial. There has been some recent progress using deep learning-based methods to accelerate qMRI \cite{zhu2023physics,cai2018single,cohen2018mr,luu2021qmtnet,liu2019mantis,liu2020high,TIAN2020117017,feng2021modl,jun2021deep,liu2021magnetic}, including MANTIS by Liu et al. \cite{liu2019mantis,liu2020high}, DeepDTI by Tian et al. \cite{TIAN2020117017}, MoDL-QSM by Feng et al. \cite{feng2021modl}, and DOPAMINE by Jun et al. \cite{jun2021deep}, all of which use supervised learning to enable rapid MR parameter mapping from undersampled k-space data.

Unsupervised or self-supervised learning methods have recently gained increasing attention due to their nature of reduced data demand. In unsupervised learning methods, the algorithm can identify patterns and structures of the input data without explicit reference pair. Self-supervised learning methods involve training a model to learn the properties and features of the input data through self-instruction. Liu et al. recently \cite{liu2021magnetic} proposed a self-supervised learning framework that incorporates an end-to-end CNN mapping and an MRI physics model to guide the generation of MR parameter maps. This method, referred to as REference-free LAtent map eXtraction (RELAX), has shown excellent performance in reconstructing undersampled qMRI data for $T_1$ and spin-spin relaxation time ($T_2$) mapping in the brain and knee. Notably, RELAX only trains on undersampled k-space datasets without the need for fully sampled k-space for reference, greatly alleviating the data limitation. The end-to-end CNN mapping in RELAX is critical in converting the undersampled k-space data into MR parameter maps through cross-domain learning, despite the end-to-end mapping (e.g., deep U-Net) requires high computing resources such as GPU memory,  needs diverse training datasets to provide sufficient image features and lacks convergence guarantees due to its nonlinear network operations and deep structures. 

Optimization algorithms can be used to unroll deep neural networks and have been applied for MRI reconstruction \cite{liang2020deep,monga2021algorithm,mccann2017convolutional}. Those algorithms stem from classic mathematical methods to solve constrained optimization problems and often have well-defined solutions. For example, Variational Network (VN) \cite{hammernik2018learning} unrolls a variational model by iteratively updating along the gradient descent direction. ADMM-net \cite{sun2016deep} learns to unroll the alternating direction method of multipliers (ADMM). ISTA-net \cite{zhang2018ista} unrolls  iterative shrinkage thresholding algorithm (ISTA) to solve the reconstruction problem with $\ell_1$ regularizations. PD-net \cite{adler2018learned} iterates the learnable primal-dual hybrid gradient algorithm (PDHG) and learns the primal and dual variables in an alternative fashion.

This paper proposes a new deep learning method for qMRI reconstruction by marrying self-supervised learning and an optimization algorithm to unroll the learning process. This proposed method is an extension of RELAX \cite{liu2021magnetic}, thus referred to as  RELAX with MOdel REinforcement (RELAX-MORE), which jointly enforces the optimization and MR physics models for rapid MR parameter mapping. RELAX-MORE aims to enable efficient, accurate, and robust MR parameter estimation while achieving subject-specific learning, featuring training on single-subject data for further addressing the data limitation. 

The rest of the paper is organized as follows: Section \ref{theory} introduces the theory and a bi-level optimization scheme of RELAX-MORE. Section \ref{method} details the implementation of lower level and upper level optimization and describes the experiment setup. Section \ref{results} provides the results and discussion. Section \ref{conclusion} concludes the paper.

\section{Theory}\label{theory}
\subsection{Supervised Learning and Self-supervised Learning for Accelerated MRI Reconstruction}
Let $ \xrm \in\ \CC^{N_x\times N_y\times N_z\times N_c}$ be the measured MR image with a resolution $N_x\times N_y\times N_z$, where $x, y, z$ are spatial dimensions and $N_c$
denotes the number of coil elements in a typical multi-coil MRI reconstruction scenario. The corresponding undersampled k-space measurement ${\frm  \in \CC }^{N_{kx}\times N_{ky}\times N_{kz}\times N_c} $  and $\frm = \Ubf \F \Cbf \xrm + n$ incoproates the MRI encoding process, where $\Ubf$ is an undersampling mask for accelerated acquisition, $\F$ is the discrete Fourier transform, $\Cbf$ denotes the coil sensitivity map and $n\in\ \mathbb{C}^{N_{kx}\times N_{ky}\times N_{kz}\times N_c}$ is the signal noise during data acquisition, and $kx, ky, kz$ are k-space dimensions.

Supervised learning uses the fully sampled image $\xrm^\ast\ \in\ \CC^{N_x\times N_y\times N_z\times N_c}$ as reference and the goal is to restore the image content from undersampled k-space data by learning network parameters $\Theta$ through minimizing the following loss function:
\begin{equation}\label{supervise_loss}
\min_\Theta{\ell\left( \xrm \left( \frm \middle| \Theta \right),\xrm^\ast\right)}.
\end{equation}
The loss function $\ell$ assesses the difference between the fully sampled $\xrm^\ast$ and the network output as the reconstructed image $\xrm \left( \frm \middle| \Theta \right)$. 
Once trained with a large amount of data pairs $(\frm, \xrm ^\ast)$, 
the network can learn the mapping between undersampled measurements $\frm$ and fully sampled data $ \xrm^\ast$ for further inference.

In self-supervised learning, the network is trained to learn the mapping from undersampled k-space data $\frm$ to itself. However, the MR physics model for image formation must be incorporated into the learning pipeline to guide the training process. The loss function in Equation \eqref{supervise_loss} can be reformulated as:
\begin{equation}\label{selfsupervise_loss}
\min_\Theta{\ell\left(  \Ubf \F \Cbf \xrm \left( \frm \middle| \Theta \right), \frm\right)}.
\end{equation}
Notably, as minimizing Equation \eqref{selfsupervise_loss} involves no fully sampled data and solely relies on learning the intrinsic structure and properties of the undersampled data itself, making it a powerful concept for  deep learning methods with limited data \cite{yaman2022zero,akccakaya2019scan,yaman2020self}. 

\subsection{Self-supervised Learning for Accelerated qMRI Reconstruction}
To extend from MRI reconstruction to qMRI reconstruction, the MR parameter maps can be denoted as $\Delta=\left\{\delta_i\right\}_{i=1}^N$ where $\delta_i$ represents each MR parameter and $N$ is the total number of MR parameters to be estimated. Given the MR signal model $\Mbf$, which is a function of $\Delta$, and the fully sampled image $\xrm ^\ast$, the MR parameters $\Delta$ can be estimated by minimizing the following problem:
\begin{equation}\label{min_Delta}
    \min_\Delta{ \ell\left( \Mbf (\Delta), \xrm^\ast\right)}.
\end{equation}
This is also referred to as model-based qMRI reconstruction. Here, the MR image $\xrm \in \CC^{N_x\times N_y\times N_z\times N_c\times N_k}$  and its undersampled k-space data $\frm \in\ \mathbb{C}^{N_{kx}\times N_{ky}\times N_{kz}\times N_c\times N_k}$ need to include $N_k$ measurements at varying imaging parameters dependent on individual qMRI method. The Equation \eqref{min_Delta} can be formulated as:
\begin{equation}\label{min_Delta_f}
\min_\Delta{\ell \left( \Ubf \F \Cbf  \Mbf \left(\Delta\right), \frm \right)},
\end{equation}
which incorporates the MRI encoding operators.

The above minimization problem \eqref{min_Delta_f} can be solved in self-supervised learning by minimizing the following loss function:
\begin{equation}\label{min_Theta}
\min_\Theta{\ell \left( \Ubf \F \Cbf \Mbf \left(\Delta \left( \frm \middle|\Theta\right)\right),\frm\right)  },
\end{equation}
where the quantitative MR parameter $\Delta\left( \frm \middle|\Theta \right)$ are parametrized from deep neural networks with learnable network parameters. In RELAX, minimizing Equation \eqref{min_Theta} with respect to trainable network parameters $\Theta$ is to train a deep neural network via network backpropagation for an end-to-end mapping network.

\subsection{The RELAX-MORE Mechanism}
The proposed RELAX-MORE extends the RELAX framework by further considering unrolling the mapping network using an optimization algorithm. This new method is designed to solve a bi-level optimization problem, where the mapping network learning and quantitative MR parameter estimation can be jointly reformulated as:
\begin{subequations} \label{bilevel}
\begin{align} 
 \min_{\Theta}  \,\,  & \ell(\Ubf \F \Cbf \Mbf (\Delta \left( \frm \middle|\Theta\right) ), \frm ) \quad \text{s.t.} \label{up}\\
& \Delta( \frm | \Theta ) 
  = \argmin_{\Delta} \phi (\Delta), \label{low} \\
\text{ where } \,\, & \phi \left(\Delta\right):=  \tfrac{1}{2}\parallel \Ubf \F \Cbf \Mbf \left(\Delta\right)- \frm \parallel_2^2+\lambda \R \left(\Delta\right). \label{phi} 
\end{align}
\end{subequations}
Like RELAX, the upper level \eqref{up} is a backward training process to optimize network parameters $\Theta$ by minimizing loss function $\ell$ via normal network backpropagation. However, unlike RELAX, the lower level \eqref{low} can be implemented as a forward process to solve for the optimal solution of $\min\limits_{\Delta}{\phi\left( \Delta \right)} $ via a gradient descent based method. The objective function $\phi$ is a regularized least square model for quantitative MR parameter optimization. The weighted regularization $\lambda \R$ provides prior information pertinent to the desired MR parameter maps. Hand-crafted regularizations can be employed, such as total variation (TV) ( $\parallel\cdot\parallel_{TV}$ ) or $\ell_1$ norm ($\parallel\cdot\parallel_1$) to promote different aspects of image features. Recent studies show that regularization $\mathcal{R}_\Theta$ using CNNs can improve feature characterization, suppressing the hand-crafted priors \cite{aggarwal2018modl,diamond2017unrolled}. 
The weight $\lambda$ can be learned and integrated into $\mathcal{R}_\Theta$. The objective function $\phi_\Theta$ depends on the network parameters $\Theta$ that learned from regularizers, which can be parametrized as the summation of CNNs with $\ell_{2,1}$ norm:
\begin{equation} \label{reg}
    \R_{\Theta}(\Delta) := \sum^N_{i=1} \| \D_{\Theta_i}(\delta_i)\|_{2,1}.
\end{equation}
Each ${\mathcal{D}_\Theta}_i$ learns to extract the prior features of different MR parameters $\delta_i$, for each   $i=1,\ldots,N$. Therefore, the RELAX-MORE for qMRI reconstruction can be mathematically described as solving a CNN regularized bi-level optimization problem:
\begin{subequations} \label{relaxmore}
\begin{align} 
 \min_{\Theta}  \,\,  & \ell(\Ubf \F \Cbf \Mbf (\Delta \left( \frm \middle|\Theta\right) ), \frm ) \quad \text{s.t.} \label{rm_up}\\
& \Delta( \frm | \Theta ) 
  = \argmin_{\Delta}  \phi_\Theta(\Delta), \notag \\
= & \argmin_{\Delta}  \HH(\Delta) + \R_\Theta(\Delta), \label{rm_low}
\end{align}
\end{subequations}
where data fidelity term $\HH(\Delta) = \tfrac{1}{2}\parallel \Ubf \F \Cbf \Mbf \left(\Delta\right)- \frm \parallel_2^2$.

\section{Methods}\label{method}
\subsection{The Lower Level Optimization: Forward Unrolling of A Learnable Descent Algorithm}
Proximal gradient descent, a widely used iterative algorithm, is implemented to unroll the lower level optimization \eqref{rm_low} in RELAX-MORE. For each unrolled iterative phase $t=1,\ldots,T$,  the following steps is used to update MR parameter $\delta_i$ for each $i=1,\ldots,N$ :
\begin{subequations}\label{prox_gd}
\begin{align} 
\bar{\delta}_i^{(t)} & = \delta_i^{(t-1)}-\alpha^{(t)}_i \nabla \tfrac{1}{2}\parallel 
 \Ubf \F \Cbf \Mbf(\{{\delta}_i^{(t-1)}\}_{i=1}^N)- \frm \parallel_2^2  \label{gd} \\
\delta_i^{(t)} &=  {\rm prox}_{\R_\Theta,\rho_t}( {\bar{\delta}}_i^{(t)} ),  \label{prox} 
\end{align}
\end{subequations}
where the proximal operator for regularization $\lambda\R(z)$ is defined as: 
\begin{equation}\label{def_prox}
    {\rm prox}_{\lambda\R,\rho}(y) := \argmin_{z} \tfrac{\rho}{2}  \parallel z-y \parallel + \lambda\R(z).
\end{equation}
This step involves implementing the proximal operation for regularization $\R$, which is equivalent to finding the maximum-a-posteriori solution for the Gaussian denoising problem at a noise level $\sqrt{\lambda/\rho}$ \cite{heide2014flexisp,venkatakrishnan2013plug}, thus the proximal operator can be interpreted as a Gaussian denoiser. However, because the proximal operator ${\rm prox}_{\mathcal{R}_\Theta,\rho_t}$ in the objective function \eqref{rm_low} does not admit closed form solution, a CNN is used to substitute ${\rm prox}_{\mathcal{R}_\Theta,\rho_t}$, where the network is constructed as a residual learning network $\G_\Theta$. Network $\G_\Theta$ is composed with CNN $\D_\Theta$, which reflects Equation \eqref{reg}, and its adjoint $\widetilde{\D}_{\Theta}$ with symmetric but reversed architecture to increase the network capability, and a soft shrinkage operator $\Scal_{\beta}$ is applied in between.  All the substitutions are mathematically proven to be effective by \cite{zhang2018ista,bian2020deep}. Then step \eqref{prox} becomes
\begin{subequations}
\begin{align}
\delta_i^{(t)} & = \G_{\Theta_i}(\bar{\delta}_i^{(t)}) + \bar{\delta}_i^{(t)}, \\
  & = \widetilde{\D}_{\Theta_i} \circ \Scal_{\beta^{(t)}_i} \circ \D_{\Theta_i} (\bar{\delta}_i^{(t)}) + \bar{\delta}_i^{(t)}, \,\, \forall i = 1,\ldots,N,
\end{align}
\end{subequations}
where the soft thresholding operator is $ \Scal_{\beta^{(t)}} = {\rm prox}_{\parallel \cdot \parallel_{2,1},\beta^{(t)}}(y) = [ {\rm sign}(y_i) \max(|y_i| - \beta^{(t)}, 0)  ] $
for vector $y = (y_1,\ldots,y_n ) \in \RR^n$ and $\beta^{(t)}$ is a soft thresholding parameter that is updated at each  phase $t=1,\ldots,T$. In summary, the forward learnable proximal gradient descent algorithm to achieve unrolled lower level optimization in RELAX-MORE can be detailed in Algorithm:
\begin{algorithm}[htbp]
\caption{\textbf{Forward Learnable Descent Algorithm for solving \eqref{rm_low}} }
\label{alg:lda}
\textbf{Input:} $\delta_i^{(0)}, \alpha_i^{(0)}, \beta_i^{(0)},  \, i = 1, \cdots, N$\\
\For{$t=1$  \KwTo $T$}
{\For{$i=1$  \KwTo $N$}
{
$\bar{\delta}_i^{(t)}  = \delta_i^{(t-1)} - $ \label{T1} 
\begin{flushright}
 $ \alpha_i^{(t)} \nabla \frac{1}{2}  \| \Ubf \F \Cbf  \Mbf(\{\delta_i^{(t-1)} \}_{i=1}^N)  - \frm \|_2^2, $\\
 \end{flushright}
$\delta_i^{(t)} = \widetilde{\D}_{\Theta_i} \circ \Scal_{\beta^{(t)}_i} \circ \D_{\Theta_i} (\bar{\delta}_i^{(t)}) + \bar{\delta}_i^{(t)} $\\
 }
 $\xrm^{(t)} = \Mbf( \{\delta_i^{(t)} \}_{i=1}^N),\,\, $\\
 } 
\textbf{output} $ \{\delta_{i}^{(T)}\}_{i=1}^N$ \textbf{and} $ \xrm^{(t)},   \forall t \in \{1,...,T\}$.  \label{lda_end}
\end{algorithm}

The soft thresholding parameter $\beta^{(t)}$ and step size $\alpha^{(t)}$ are learnable and updating during each phase. The final outputs of the algorithm are estimated intermediate MR images $\xbf^{(t)}, \forall t = 1,\ldots,T$ for all iterative phases and the reconstructed MR parameter maps  $ \delta_{i}^{(T)}, \forall i = 1,\ldots,N$ of the last phase.

\subsection{Initial Input of the Algorithm}
Using an initial MR parameter input $\delta_{i}^{(0)}$  that is closer to the optimal solution of the Algorithm can lead to better results and eliminate the need for excessive iterations in gradient descent. To achieve a favorable input, initialization networks are used to learn $\delta_{i}^{(0)}, \forall i=1,…,N$ from undersampled k-space data. First, the undersampled k-space data is converted into the image domain, followed by a Coil Combination Network to combine all coil-wise images.
$N_k$ coil-combined images are produced to represent $N_k$ measurements during qMRI acquisition. Next, all images are concatenated together as input into the $\delta$-initialization Network to obtain $\delta_{i}^{(0)}, \forall i=1,…,N$. The initial MR images can also be obtained from the signal model using $ \xrm^{(0)} = \Mbf( \{ \delta_i^{(0)} \}_{i=1}^N)$, which will be subsequently used in the loss function $\ell$ for training the overall network framework in upper level optimization in Equation \eqref{rm_up}.

\subsection{The Upper Level Optimization: Backward Training of Overall Network}
\begin{figure*}[t]
\centering
\centerline{\includegraphics[width=0.82\paperwidth]{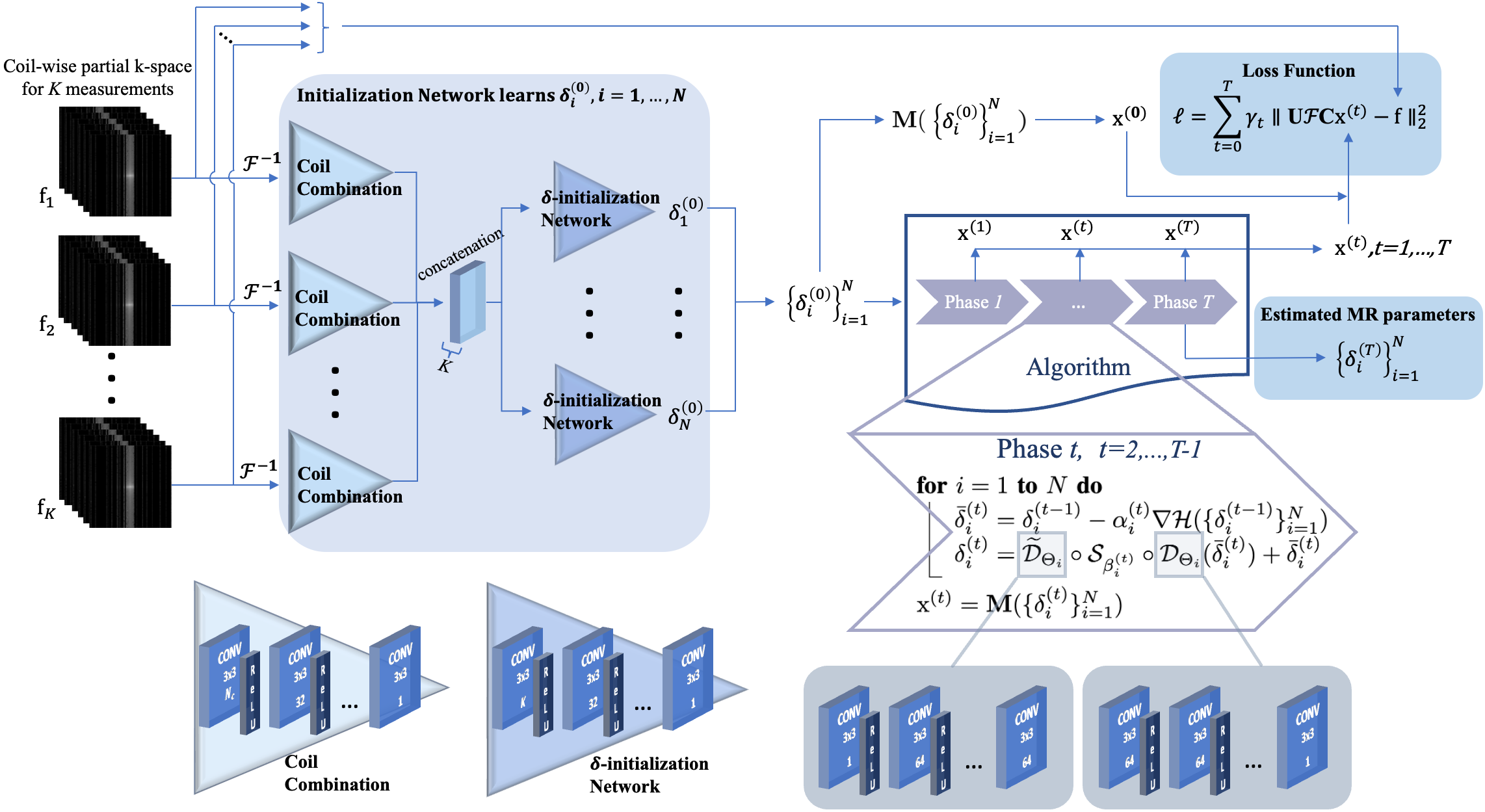}}
\caption{Schematic framework for implementing RELAX-MORE. }
\label{fig:framework}
\end{figure*}

The loss function \eqref{rm_up} for RELAX-MORE is a summation of the weighted mean squared error (MSE) between undersampled k-space measurement $\frm$  and the k-space representation of estimated MR images $\xrm^{(t)}, \forall t=0,\ldots,T$ for all unrolled phases. Therefore, the overall network training objective in the upper level optimization is to minimize the loss function $\ell$ explicitly expressed as 
\begin{equation}\label{loss}
\ell(\Ubf \F \Cbf  \xrm^{(t)}, \frm ) = \sum^T_{t=0} \gamma_t \parallel  \Ubf \F \Cbf  \xrm^{(t)} - \frm \parallel^2_2, 
\end{equation}
where the $\gamma_t$ are the weighting factors for the MSE in order of the importance of the output from each phase and the initial $\xrm^{(0)}$.

The detailed framework of the proposed RELAX-MORE is illustrated in Fig. \ref{fig:framework} The Coil combination Network and $\delta$-initialization Network applied 4 repeats of Conv($3\times3$ kernel size, 32 kernels) followed by a ReLU activation function. The last convolution layer has a single kernel Conv($3\times3$, 1). The CNN $\D$, applied 8 Conv($3\times3$, 64) with a ReLU and the last convolution also has a single kernel Conv($3\times3$, 1). The CNN $\widetilde{\D}$ applied a reversed network structure.

\subsection{Experiment Setup}\label{setup}
To investigate the feasibility of RELAX-MORE method for reconstructing accelerated qMRI, we used the widely studied $T_1$ mapping through the vFA method \cite{wang1987optimizing} as an example. Herein, the MR signal model $\Mbf$ can be expressed as:
\begin{equation}\label{vfa}
    \Mbf_{k}(\Tbf_1, \Ibf_0) = \Ibf_0 \cdot \frac{(1-e^{-TR/ \Tbf_1}) \sin{\eta_k}}{ 1 -e^{-TR/\Tbf_1} \cos{\eta_k}}, 
\end{equation}
where several MR images are acquired at multiple flip angles $\eta_k$ for $k = 1,\ldots, N_k $.  $ \Tbf_1  \in \RR^{N_x \times N_y \times N_z }, \Ibf_0 \in \CC^{N_x \times N_y \times N_z } $ are the spin-lattice relaxation time map and proton density map,  respectively, reflecting the imaged tissue relaxation properties, with $T_1$ value sensitive to many brain and knee joint diseases. The set of MR parameters estimated in this model are $\Delta = \{ \Tbf_1,\Ibf_0\}$. The flip angles and other imaging parameters, such as repetition time TR are pre-determined.

The experiments include in-vivo studies on the brain and knee of healthy volunteers and ex-vivo phantom studies, all of which were carried out on a Siemens 3T Prisma scanner. For the brain study, the purpose is to investigate the efficiency and performance of RELAX-MORE and compare it with other state-of-the-art reconstruction methods. The vFA on the brain of five subjects was performed in the sagittal plane using a spoiled gradient echo sequence at imaging parameters TE/TR = $12/40$ ms, FA = $5^\circ, 10^\circ, 20^\circ, 40^\circ$, FOV = $230 \times 230 \times 160$ mm and matrix size = $176 \times 176 \times 48$ with a dedicated 20-channel head coil.
For the knee study, the purpose is to investigate the robustness of RELAX-MORE against severe noise contamination and image imperfection. Therefore, the vFA on one knee was performed using a 4-channel receiving-only flex coil at parameters TE/TR = $12/70$ ms, FA = $10^\circ, 20^\circ, 40^\circ$, FOV = $160\times137\times108$ mm and matrix size $224\times192\times36$. For the phantom study, the purpose is to investigate the reconstruction accuracy of RELAX-MORE. The vFA phantom data was acquired along the coronal plane with sequence parameters TE/TR = $12/80$ ms, FA = $5^\circ, 10^\circ, 20^\circ, 40^\circ, 60^\circ$, FOV = $170\times170\times60$ mm and matrix size $128\times128\times12$ using the 20-channel head coil. 
In all 3 studies, to minimize bias on $T_1$ estimation due to $B_1^+$ inhomogeneity and imperfect spoiling, $B_1^+$ maps were separately acquired for compensation \cite{sacolick2010b1}, and $169^{\circ}$ linear RF phase increments between subsequent repetition cycles and strong gradient spoilers were applied to minimize the impact of imperfect spoiling \cite{preibisch2009influence}. 

\begin{figure}[t]
\centering
\centerline{\includegraphics[width=0.4\paperwidth]{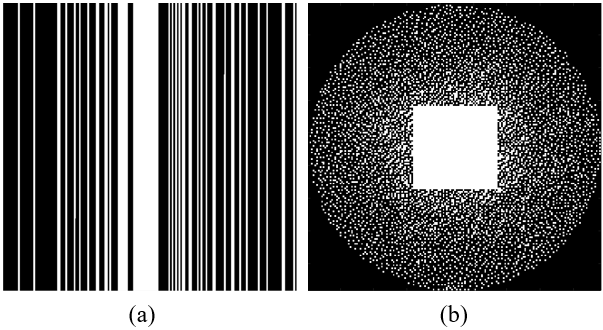}}
\caption{Exemplified masks: (a) 3x 1D Cartesian variable density undersampling mask and (b) 4x 2D Poisson disk undersampling mask.}
\label{fig:mask}
\end{figure}
The fully acquired MRI k-space data were undersampled retrospectively using two undersampling schemes: 1) 1D Cartesian variable density undersampling at acceleration factor AF = $3\times$ with the 16 central k-space lines fully sampled (Fig. \ref{fig:mask}(a)) and 2) 2D Poisson disk undersampling at AF = $4\times$ with the central $51\times51$ k-space portion fully sampled (Fig. \ref{fig:mask}(b)). The undersampling patterns were varied for each flip angle, like in previous studies \cite{liu2019mantis,liu2021magnetic}. The coil sensitivity maps were estimated from ESPIRiT \cite{uecker2014espirit}.

The selection of network hyperparameters was non-trivial and optimized by empirically searching on a reasonably large range. More specifically, we used initial $\alpha_i^{(0)}=0.1, \beta_i^{(0)}=1e^{-5}$ for brain images; initial $\alpha_i^{(0)}=0.5, \beta_i^{(0)}=1e^{-7}$ for knee images; and $\alpha_i^{(0)}=0.9, \beta_i^{(0)}=1e^{-3}$ for phantom images. The loss function applied $\gamma_t=1$ at $t=T$ and $\gamma_t=1e^{-4}$ for other phase steps. The network parameters $\Theta$ were initialized using Xavier initialization \cite{glorot2010understanding}. Because RELAX-MORE trains on single-subject data, the batch size was set to include the entire image volume. The network was trained for $10,000$ epochs using the Adam optimizer \cite{kingma2014adam}. The learning rate, which controls the step size taken in each iteration of optimizing the loss function, was set to $1e^{-4}$ for brain and knee images, and $5e^{-4}$  for the phantom image. It should be noted that, unlike conventional deep learning reconstruction methods where training and testing are two separate steps, in RELAX-MORE with subject-specific self-supervised learning, the reconstruction has readily concluded once the training has converged for one subject. All the programming in this study was implemented using Python language and PyTorch package, and experiments were conducted on one NVIDIA A100 80GB GPU and an Intel Xeon 6338 CPU at Centos Linux system.

\section{Results}\label{results}
\subsection{Impact of Unrolling Gradient Descent Algorithm}\label{impact_alg}
\begin{figure*}[t]
\centering
\centerline{\includegraphics[width=0.82\paperwidth]{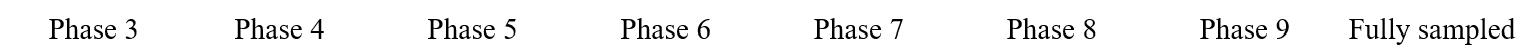}}
\centerline{\includegraphics[width=0.82\paperwidth]{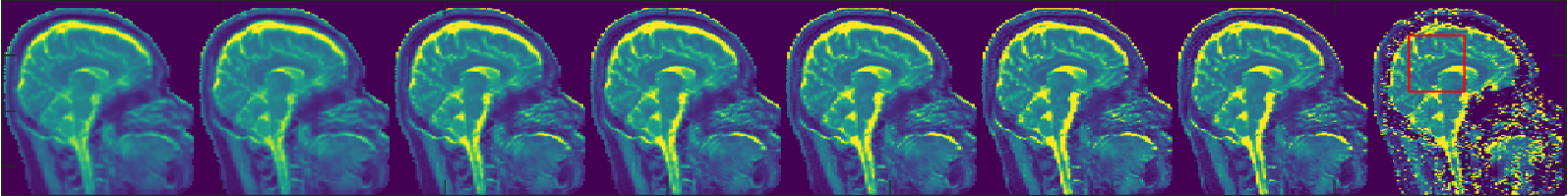}}
\centerline{\includegraphics[width=0.82\paperwidth]{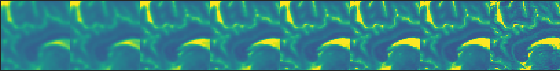}}
\caption{$T_1$ maps generated from RELAX-MORE at different phases of the forward learnable descent algorithm $(3 \le T \le 9 )$ and $T_1$ map obtained from fully sampled data using the standard pixel-wise fitting. While performance gain is significant with increasing phases at lower phase numbers, incremental performance gain is observed after phase 7.  }
\label{fig:phases}
\end{figure*}
The reconstruction performance of RELAX-MORE is affected by the degree of unrolling the gradient descent algorithm. Fig. \ref{fig:phases} demonstrates the evolution of the estimated $T_1$ maps from fully sampled data at different total unrolling phase numbers from $T=3$ to $T=9$ of the gradient descent algorithm. The larger $T$ reflects deeper unrolling operation and thus requires more computing resources. Referring to the zoom-in views of Fig. \ref{fig:phases}, one can observe that with increasing total phase number, the algorithm starts to better distinguish the $T_1$ values from white matter (WM), grey matter (GM), and cerebrospinal fluid (CSF) regions. Initially blurry at a lower $T$, the tissue details start to sharpen with increasing $T$, resembling better the fully sampled $T_1$ map obtained using the standard pixel-wise fitting. However, the differences between phase $T = 7, 8$ and $9$ are negligible, indicating that performance gain can reach a plateau with a sufficient depth of unrolling steps. Therefore, to ensure consistent experiment setup and balance the trade-off between algorithm complexity and reconstruction performance, the number of $T$ was set to $8$ for all the experiments thereafter. This result illustrates the effectiveness of unrolling gradient descent algorithm in the RELAX-MORE framework.

\subsection{Comparison with State-of-the-Art Methods}\label{compare}
RELAX-MORE was compared with two state-of-the-art non-deep learning qMRI reconstruction methods and one self-supervised deep learning method. These methods include 1) Locally Low Rank (LLR) \cite{zhang2015accelerating}, where image reconstruction was first performed, followed by pixel-wise parameter fitting. LLR exploits the rank-deficiency of local image regions along the acquisition parameter dimension to accelerate parameter mapping; 2) Model-TGV \cite{maier2019rapid}, a model-based qMRI reconstruction method that improves the piece-wise constant region restriction of total variation through a generalization of the total variation theory and 3) RELAX \cite{liu2021magnetic}, an end-to-end self-supervised deep learning method for rapid MR parameter mapping. LLR and Model-TGV were implemented using the recommended optimization parameters in their original papers and code. Contrary to the original RELAX implementation, which was trained using many subjects, in the interest of a fair comparison, our implementation of RELAX was carried out on a single-subject training using an Adam optimizer with $10,000$ epochs.

\begin{figure*}[htbp]
\centering
\centerline{\includegraphics[width=0.85\paperwidth]{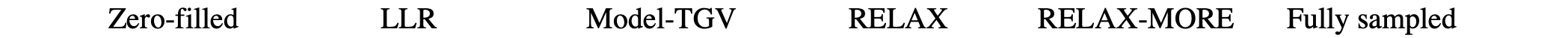}}
\centerline{\includegraphics[width=0.85\paperwidth]{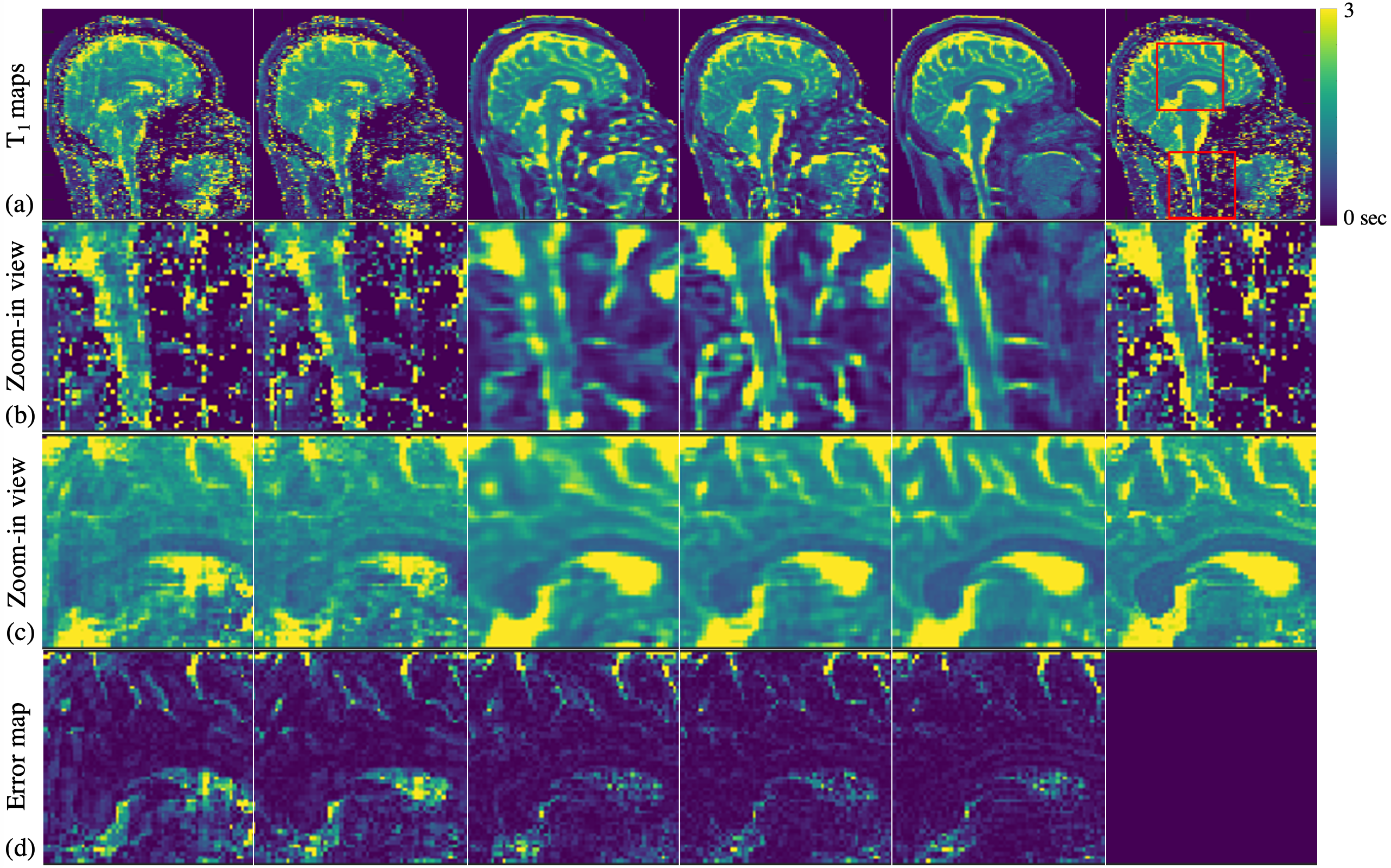}}
\caption{(a) Qualitative comparison among different methods using $3\times$ 1D Cartesian variable density undersampling mask;  (b) Zoom-in view with details of the region outside the brain; (c) Zoom-in view with in-brain detail; (d) pixel-wise error maps of the in-brain region details.}
\label{fig:ar3}
\end{figure*}
\begin{figure*}[htbp]
\centering
\centerline{\includegraphics[width=0.85\paperwidth]{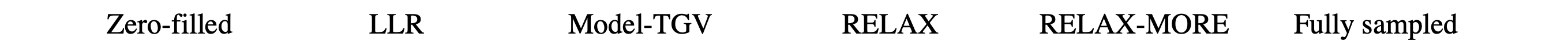}}
\centerline{\includegraphics[width=0.85\paperwidth]{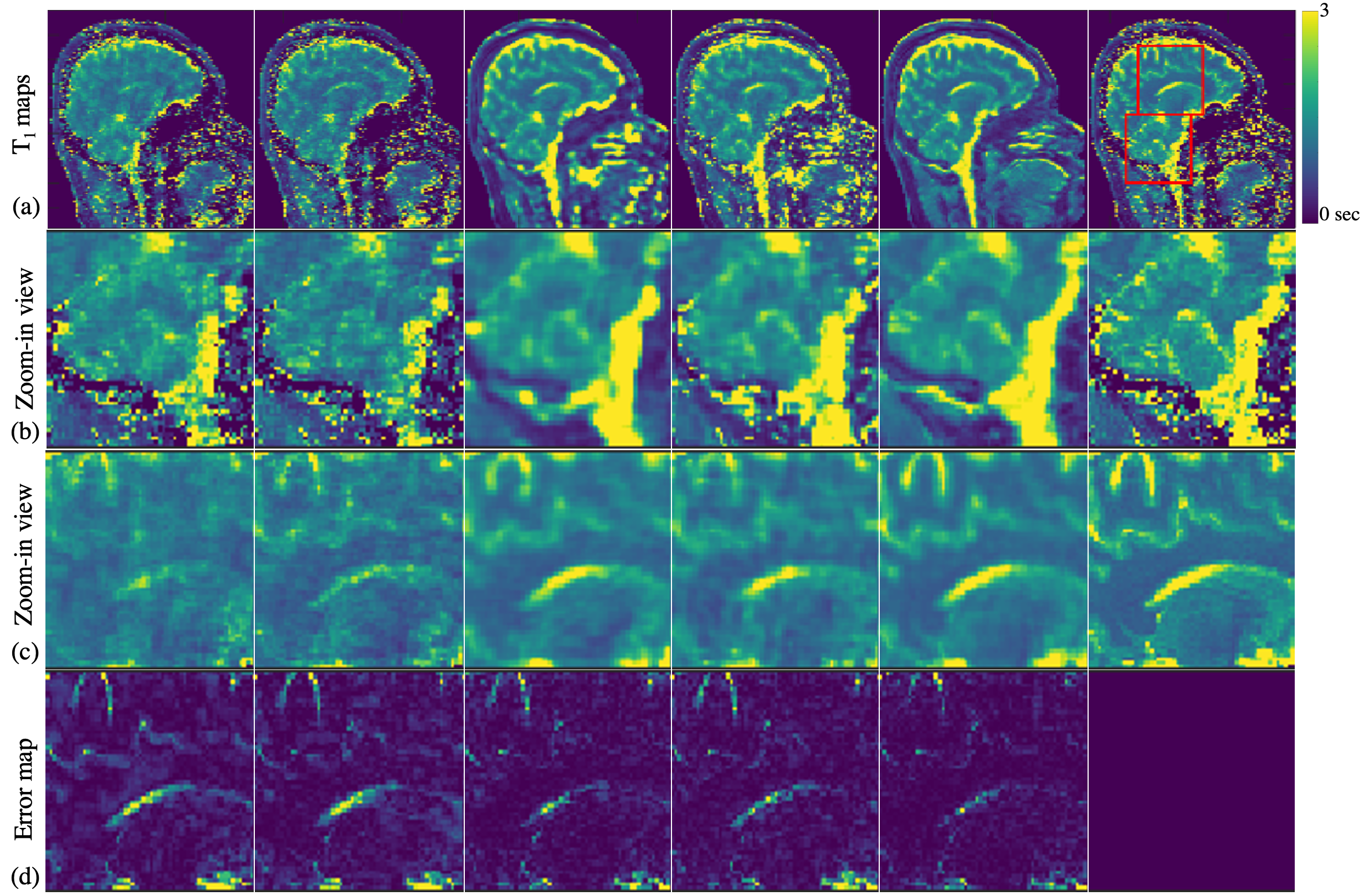}}
\caption{(a) Qualitative comparison among different methods using $4\times$ 2D Poisson disk undersampling mask;  (b) Zoom-in view with details of the cerebellum; (c) Zoom-in view with in-brain detail; (d) pixel-wise error maps of the in-brain region details.}
\label{fig:ar4}
\end{figure*}
$T_1$ maps estimated from $3\times$ 1D Cartesian variable density undersampling using the different methods are presented in Fig. \ref{fig:ar3}. As can be observed in Fig. \ref{fig:ar3}(a), the zero-filled $T_1$ map obtained by pixel-wise fitting the undersampled images is noisy, displaying ripple artifacts due to aliasing. Although LLR can partially remove these artifacts, it retains the noisy signature. Model-TGV averages out the noise, producing a cleaner tissue appearance and better $T_1$ contrast, but it is over-smoothed, resulting in blurry maps. On the other hand, RELAX removes noises and artifacts and produces sharpened maps but remains somewhat blocky. This is hypothesized due to the difficulty of converging the end-to-end network using single-subject data. RELAX-MORE produces artifact-free $T_1$ maps with anticipated excellent performance for noise removal, resulting in a good performance in appearance and contrast. This is further witnessed in the zoom-in view of Fig. \ref{fig:ar3}(b), where the zero-filled, LLR, and even fully sampled images cannot reliably estimate $T_1$ in the cervical spine where the signal is low due to insufficient head coil coverage. Those pixel-wise fitting methods are typically prone to image noise contamination. Although Model-TGV and RELAX can better estimate $T_1$ in these low SNR regions, the blurry Model-TGV results make it difficult to distinguish between the spinal cord and subarachnoid space, whereas RELAX shows a disconnect in the subarachnoid space which may be due to over-sharpening. However, RELAX-MORE shows remarkably  good performance in maintaining contrast and tissue details at those regions, enabling clear distinction between the spinal cord and subarachnoid space regions. In Fig. \ref{fig:ar3}(c) of the cerebrum zoom-in view, RELAX-MORE consistently exhibits the best reconstruction performance in correcting artifacts, removing noises, and preserving tissue contrast and details. Referring to the error maps (Fig. \ref{fig:ar3}(d)), which is the absolute difference between the estimated $T_1$ maps in Fig. \ref{fig:ar3}(c) and the fully sampled $T_1$ map, overall, zero-filled shows the most significant error followed by LLR. Model-TGV and RELAX exhibit similar error maps, whereas RELAX-MORE produces the least error.

$T_1$ maps estimated from $4\times$ 2D Poisson disk undersampling are shown in Fig. \ref{fig:ar4}. Comparing the images in Fig. \ref{fig:ar4}(a), the overall signature differences are like the $3\times$ 1D undersampling case except for the undersampling artifacts being noise-like due to 2D undersampling. Referring to the zoom-in view and comparing the estimated $T_1$ maps again show similar differences to the $3\times$ 1D undersampling case. However, compared with other methods, RELAX-MORE can clearly distinguish between WM and GM $T_1$ of the cerebellum in Fig. \ref{fig:ar4}(b), particularly in the posterior part. The absolute difference maps taken between the estimated $T_1$ maps in Fig. \ref{fig:ar4}(c) and the fully sampled map, shown in Fig. \ref{fig:ar4}(d), exhibit similar results to the $3\times$ 1D undersampling case with RELAX-MORE showing the least error. This is likely achieved through integrating deep learning capability and unrolling gradient descent algorithm with proximal operator prioritizing noise suppression without compromising the fidelity and clarity of the underlying tissue structure.

Further performance comparison was carried out using peak-signal-to-noise-ratio (PSNR), structural similarity index measure (SSIM), and normalized mean squared error (NMSE) as evaluation metrics for $T_1$ maps. PSNR is a measbure of the quality of the reconstruction, while SSIM is a measure of the similarity between two maps. PSNR, SSIM, and NMSE are defined in the following:
\begin{equation}
    PSNR(\vrm, \vrm^*) = 20\log_{10} (\max(|\vrm^*|)/\tfrac{1}{N} \parallel \vrm - \vrm^* \parallel^2),
\end{equation}
\begin{equation}
    SSIM(\vrm, \vrm^*) = \frac{(2 \mu_\vrm  \mu_{\vrm^*} +c_1)(2 \sigma_{\vrm \vrm^*} + c_2)}{(\mu_\vrm^2 +\mu_{\vrm^*}^2 + c_1)(\sigma_\vrm^2 + \sigma_{\vrm^*}^2+ c_2)},
\end{equation}
\begin{equation}
    NMSE(\vrm, \vrm^*) = \parallel \vrm^* - \vrm \parallel^2_2 / \parallel \vrm^* \parallel^2_2, 
\end{equation}
where $\vrm, \vrm^*$ represent the estimated map and reference map. $\mu_\vrm , \mu_{\vrm^*}$ are local means of pixel intensity, $ \sigma_\vrm, \sigma_{\vrm^*}$ are the standard deviations and $\sigma_{\vrm \vrm^*}$ is covariance between $\vrm$ and $\vrm^*$, $C_1 = (K_1 L)^2, C_2 = (K_2L)^2 $ are two constants that avoid denominator to be zero, and $K_1 = 0.01, K_2 = 0.03$. $L$ is the largest pixel value of the magnitude of image.

$T_1$ maps obtained from fully sampled data were used for $\vrm^*$, as a reference to compare the performance among different methods. The metric calculations were carried out on results from $3\times$ 1D Cartesian variable density undersampling, using the brain regions from all five subjects. 

\begin{figure}[htbp]
\centering
\centerline{\includegraphics[width=0.4\paperwidth]{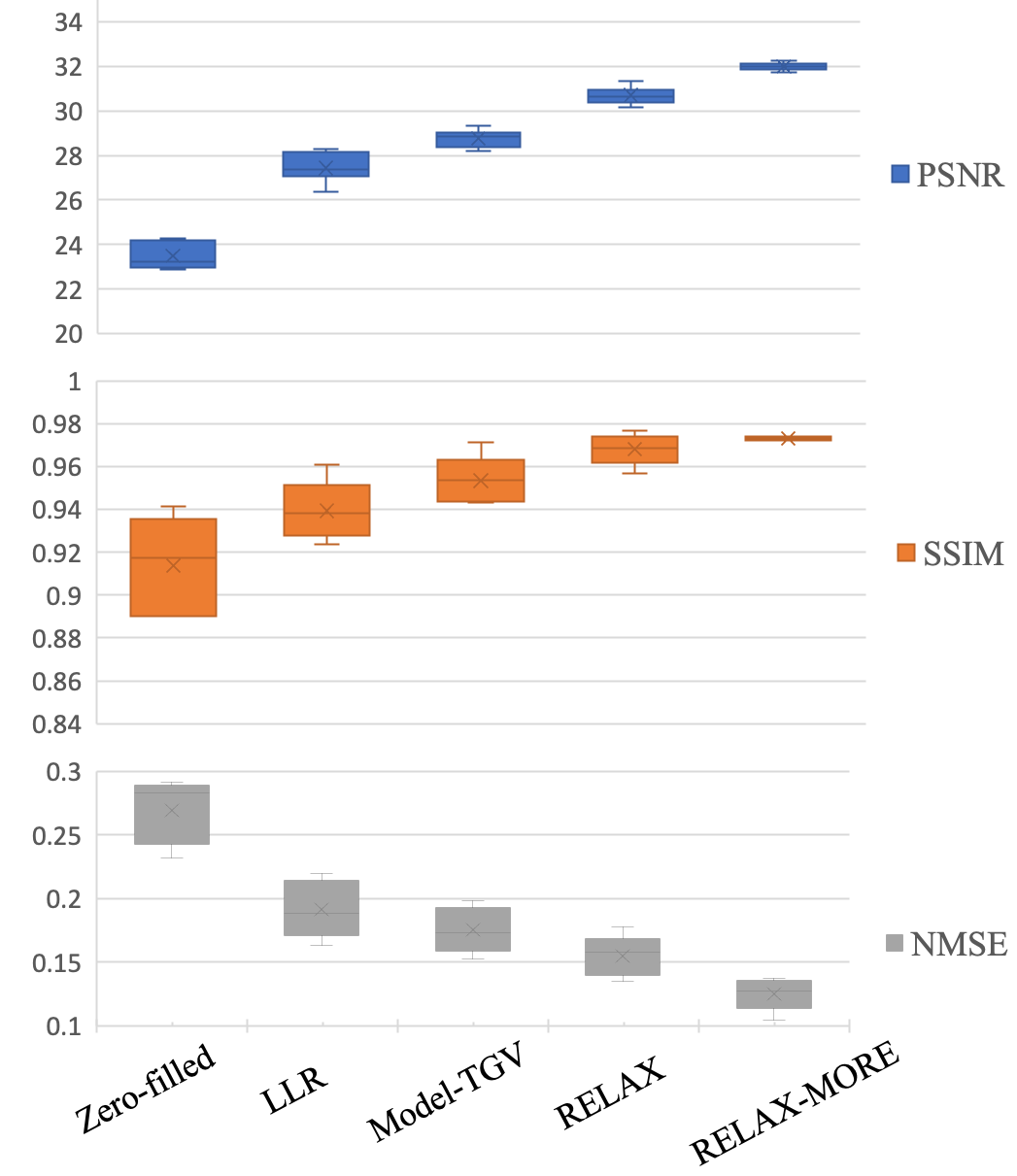}}
\caption{Quantitative comparison between the estimated $T_1$ maps with $3\times$ 1D Cartesian mask and the reference $T_1$ maps estimated from fully sampled data. The comparison was made based on three different measures, namely PSNR, SSIM, and NMSE with median line and mean markers $\times$. }
\label{fig:barchart}
\end{figure}
The analysis results are presented in Fig. \ref{fig:barchart}, where the average values of PSNR, SSIM and NMSE along with corresponding standard deviations among five subjects are presented using box-whisker plots. For both PSNR and SSIM, RELAX-MORE produces the highest mean value, followed by RELAX, Model-TGV and LLR. Both PSNR and SSIM of RELAX-MORE show the highest consistency, as evidenced by its small standard deviation compared to other methods. Zero-filled shows the lowest mean values for PSNR and SSIM with the largest standard deviation, which can be attributed to aliasing artifacts. RELAX-MORE shows the lowest mean NMSE, followed by RELAX, Model-TGV and LLR. Zero-filled shows the highest mean NMSE. Overall, in agreement with the qualitative observation in Figs. \ref{fig:ar3} and \ref{fig:ar4}, RELAX-MORE performs superiorly in terms of reconstruction fidelity, structure and texture preservation, and noise suppression, outperforming other state-of-the-art methods.

\subsection{Ablation Study: Accuracy and Robustness}
\begin{figure}[htbp]
\centering
\centerline{\includegraphics[width=0.4\paperwidth]{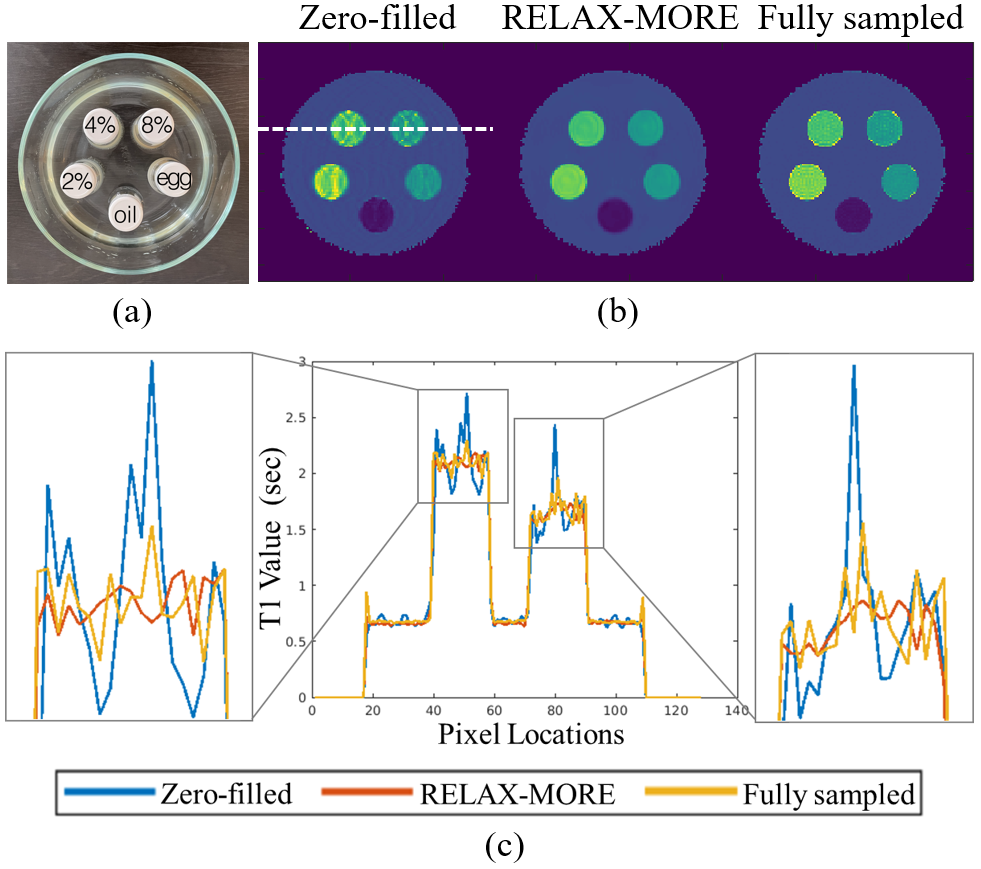}}
\caption{(a) Phantom and (b) corresponding $T_1$ maps generated from Zero-filled, RELAX-MORE, and fully sampled phantom data. Zero-filled and RELAX-MORE $T_1$ maps are estimated from data undersampled using a $3\times$ 1D Cartesian undersampling mask. (c) Line profile, indicated in (b), along with zoom-in views showing the difference in $T_1$ values. }
\label{fig:phantom}
\end{figure}
An ablation study was carried out further to evaluate the parameter estimation accuracy of the proposed method. To analyze RELAX-MORE in discrete tissue environment, it was applied to undersampled data from a phantom composed of five 20 mL vials consisting of peanut oil and $2\%, 4\%, 8\%$ Agar dissolved in DI water and boiled egg white, all immersed in a 200 $\mu$M $\rm{MnCl_2}$ water bath (Fig. \ref{fig:phantom}(a)). From left to right, Fig. \ref{fig:phantom}(b) shows the $T_1$ maps obtained from zero-filled data, RELAX-MORE and fully sampled data. Zero-filled $T_1$ map exhibits ripple artifacts due to the undersampling, whereas these artifacts are eliminated in RELAX-MORE. Comparing the line profile indicated by the line in Fig. \ref{fig:phantom}(b) and shown in Fig. \ref{fig:phantom}(c), zero-filled $T_1$ profile resembles amplitude modulation, where the periodic amplitude fluctuation stems from aliasing effects and the higher frequency fluctuation reflects noise. In good agreement with the fully sampled $T_1$ profile, RELAX-MORE’s $T_1$ profile not only removes the periodic amplitude fluctuation but also smoothens the oscillations, demonstrating its ability to average out noise, remove aliasing artifacts while maintaining high accuracy for parameter estimation in a wide range. 

\begin{figure}[htbp]
\centering
\centerline{\includegraphics[width=0.4\paperwidth]{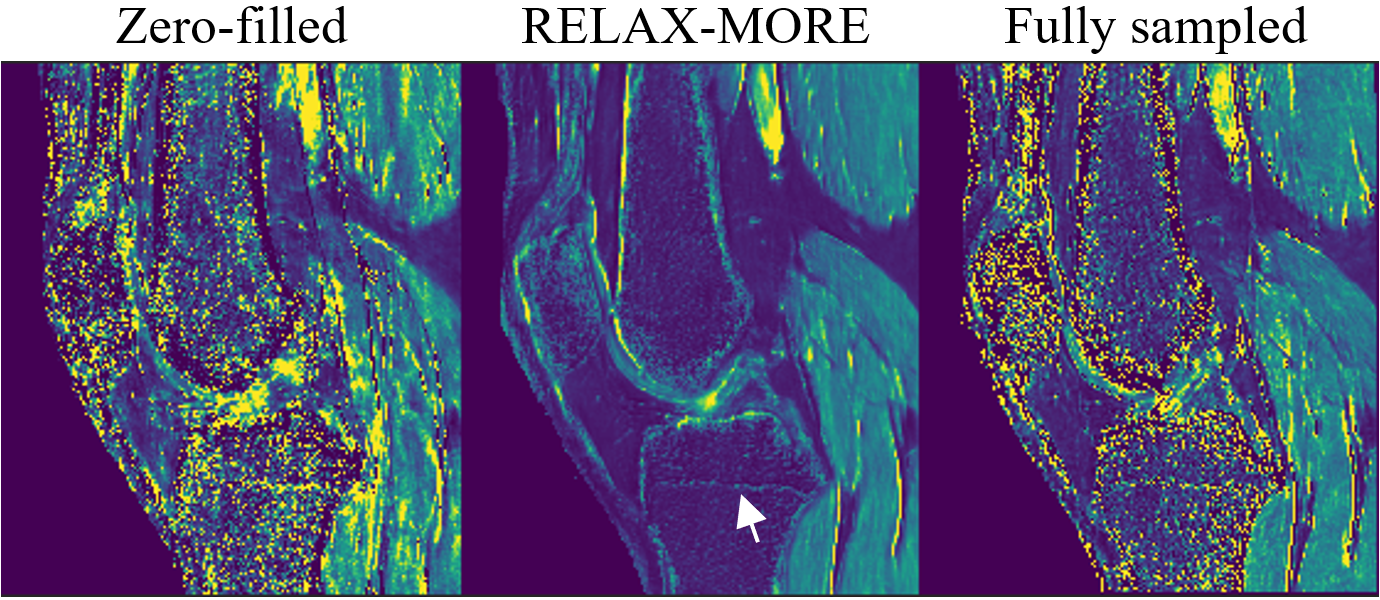}}
\caption{Estimated $T_1$ maps for knee images using $3\times$ 1D Cartesian undersampling mask. }
\label{fig:knee}
\end{figure}
The ablation study was also carried out to evaluate the parameter estimation robustness of RELAX-MORE against severe imaging imperfection and the generalizability of RELAX-MORE on different anatomies such as knee joints. Fig. \ref{fig:knee} shows the $T_1$ maps from a representative sagittal slice of the knee obtained from zero-filled, RELAX-MORE, and fully sampled data. The MRI data was purposely acquired using a 4-channel receiving-only flex coil, which provides insufficient coil coverage and a poor signal-to-noise ratio for knee MRI. As evident in Fig. \ref{fig:knee}, the fully sampled $T_1$ map presents enormous noises at bone, cartilage, ligament, meniscus, and muscle and the fine structure, such as the epiphyseal plate (highlighted by a white arrow) is contaminated by noises. The zero-filled $T_1$ map from $3\times$ 1D Cartesian acceleration also presents severe undersampling artifacts combined with the noise rendering erroneous $T_1$ quantification. However, RELAX-MORE successfully removes all the artifacts, suppresses the unwanted image noises in $T_1$ quantification, and provides a surprisingly favorable quantification of different knee joint structures. RELAX-MORE demonstrates its high robustness and generalizability for reconstructing the in-vivo brain and in-vivo knee joint, making it a widely applicable method for different anatomies.

\subsection{Transfer Learning: Computing Time Efficiency}
RELAX-MORE is a subject-specific method that utilizes self-supervised learning for efficient qMRI reconstruction. While RELAX-MORE can perform well using single-subject data, as shown in all other experiments, the reconstruction process (e.g., network training) needs to be conducted for each subject data. In this experiment, we investigated transfer learning to improve the computing time efficiency for the training/reconstruction of RELAX-MORE. Using transfer learning, the network weights after training on one brain data were then applied as the starting point to train the brain data of another subject.

\begin{figure}[htbp]
\centering
\centerline{\includegraphics[width=0.4\paperwidth]{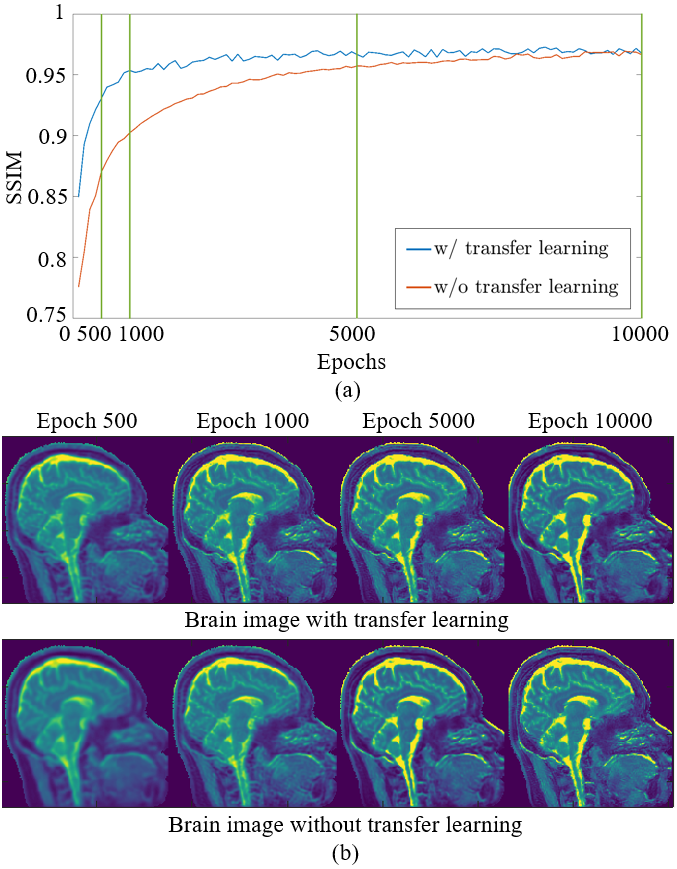}}
\caption{(a) SSIM calculated as a function of epoch for training with and without transfer learning. (b) The corresponding estimated $T_1$ maps with (top row) and without (bottom row) transfer learning at epochs 500, 1000, 5000 and 10000, indicated by vertical green lines in (a).}
\label{fig:transfer}
\end{figure}
In Fig. \ref{fig:transfer}, the performance of RELAX-MORE with and without transfer learning is presented by plotting the SSIMs for $T_1$ estimations as a function of epoch number at $3\times$ 1D Cartesian acceleration. Comparing the two SSIM curves in Fig. \ref{fig:transfer}(a), it is observed that transfer learning starts at a higher initial SSIM value during the initial training phase and increases faster compared to the case without transfer learning. This implies that transfer learning benefits the model to converge more efficiently due to commencing training with pre-trained weights. This is further confirmed by comparing the $T_1$ maps generated with (Fig. \ref{fig:transfer}(b) top row) and without (Fig. \ref{fig:transfer}(b) bottom row) transfer learning, where at low epoch values (500, 1000), transfer learning generated $T_1$ maps show better $T_1$ clarity and contrast between different tissue types compared to its non-transfer learning counterpart. As the training epoch increases, both SSIM curves flatten, indicating that the model has reached stable performance. The results suggest that transfer learning can improve the computing time efficiency of RELAX-MORE with reduced training/reconstruction time. At approximately 0.5 seconds per epoch using our GPU device for training the 3D brain data, transfer learning can reduce the reconstruction time to less than 10 min to reach good performance. With the advance of new techniques \cite{zhuang2020comprehensive},  transfer learning can be a practical approach to further improve the reconstruction timing efficiency for RELAX-MORE.

\section{Conclusion}\label{conclusion}
This paper proposes a novel self-supervised learning method, RELAX-MORE, for qMRI reconstruction.  This proposed method uses an optimization algorithm to unroll a model-based qMRI reconstruction into a deep learning framework, enabling the generation of highly accurate and robust MR parameter maps at imaging acceleration. Unlike conventional deep learning methods requiring a large amount of training data, RELAX-MORE is a subject-specific method that can be trained on single-subject data through self-supervised learning. Furthermore, in our experiments, RELAX-MORE outperforms several state-of-the-art conventional and deep learning methods for accelerated $T_1$ maps. This work also demonstrates several superior aspects of RELAX-MORE in overall performance, accuracy, robustness, generalizability, and computing timing efficiency, making this method a promising candidate for advancing accelerated qMRI for many clinical applications.

\bibliographystyle{IEEEtran}
\bibliography{main}
\end{document}